\documentclass{siamart1116}

\usepackage{algorithm}
\usepackage[algo2e,ruled,vlined,linesnumbered,commentsnumbered]{algorithm2e}
\IncMargin{1em}
\SetCommentSty{scriptsize}
\SetKwComment{tcc}{\{}{\}}
\SetKwInput{Input}{Input}
\SetKwInput{Output}{return}
\SetKwIF{If}{ElseIf}{Else}{if}{}{else if}{else}{endif}
\SetKwFor{While}{while}{}{endw}
\SetKwFor{ForEach}{for each}{}{endfch}

\usepackage{amsmath}
\usepackage{amsfonts}
\usepackage{graphicx}
\usepackage{epstopdf}
\usepackage{algorithmic}
\usepackage{hyperref}
\usepackage{booktabs}
\usepackage{siunitx}

\usepackage{tikz}
\newcommand{\tikzcircle}[2][red,fill=red]{\tikz[baseline=-0.5ex]\draw[#1,radius=#2] (0,0) circle;}

\newcommand{\TheTitle}{Scaling Structured Multigrid to 500K+ Cores through Coarse-Grid Redistribution\footnote{Los Alamos Report LA-UR-17-22886}}
\newcommand{\TheShortTitle}{Scaling Multigrid through Coarse-Grid Redistribution}
\newcommand{\TheAuthors}{A. Reisner, L. N. Olson, and J. D. Moulton}
\headers{\TheShortTitle}{\TheAuthors}

\title{{\TheTitle}}

\author{  Andrew Reisner\thanks{Department of Computer Science, University of Illinois at Urbana-Champaign.}
  \and
  Luke N. Olson\footnotemark[1].
  \and
  J. David Moulton\thanks{Applied Mathematics and Plasma Physics, Los Alamos National Laboratory.}
}

\ifpdf\hypersetup{  pdftitle={\TheTitle},
  pdfauthor={\TheAuthors}
}
\fi

\newcommand{\west}{\textnormal{w}}
\newcommand{\east}{\textnormal{e}}
\newcommand{\south}{\textnormal{s}}
\newcommand{\north}{\textnormal{n}}
\newcommand{\southwest}{\textnormal{sw}}
\newcommand{\northwest}{\textnormal{nw}}
\newcommand{\southeast}{\textnormal{se}}
\newcommand{\northeast}{\textnormal{ne}}

\begin{document}
\maketitle

\begin{abstract}
  The efficient solution of sparse, linear systems resulting from the
discretization of partial differential equations is crucial to the
performance of many physics-based simulations.  The algorithmic
optimality of multilevel approaches for common discretizations makes
them a good candidate for an efficient parallel solver.  Yet, modern
architectures for high-performance computing systems continue to
challenge the parallel scalability of multilevel solvers.  While
algebraic multigrid methods are robust for solving a variety of
problems, the increasing importance of data locality and cost of data
movement in modern architectures motivates the need to carefully
exploit structure in the problem.

Robust logically structured variational multigrid methods, such as
Black Box Multigrid (BoxMG), maintain structure throughout the
multigrid hierarchy.  This avoids indirection and increased
coarse-grid communication costs typical in parallel algebraic
multigrid.  Nevertheless, the parallel scalability of structured
multigrid is challenged by coarse-grid problems where the overhead in
communication dominates computation.  In this paper, an algorithm is
introduced for redistributing coarse-grid problems through incremental
agglomeration.  Guided by a predictive performance model, this
algorithm provides robust redistribution decisions for structured
multilevel solvers.

A two-dimensional diffusion problem is used to demonstrate the
significant gain in performance of this algorithm over the previous
approach that used agglomeration to one processor.
In addition, the parallel scalability of
this approach is demonstrated on two large-scale computing systems,
with solves on up to 500K+ cores.

\end{abstract}



\section{Introduction}
The efficient solution of large, sparse linear systems resulting from
the discretization of elliptic partial differential equations (PDEs)
is crucial to the performance of many numerical simulations. Although
there has been significant progress in developing general Algebraic
Multigrid (AMG) solvers~\cite{Trottenberg:2000}, modern
high-performance computing (HPC) architectures continue to pose
significant challenges to parallel scalability and
performance (e.g., ~\cite{Baker2012,gahvari-2013,Bienz2016}).
These challenges include reducing data movement,
increasing arithmetic intensity, and identifying opportunities to
improve resilience, and are more readily addressed in settings where
problem structure can be identified and exploited.  For example,
robust structured variational multigrid methods, such as Black Box
Multigrid (BoxMG)~\cite{Dendy1982,Dendy1983}, take advantage of
direct memory addressing and fixed stencil patterns throughout the
multigrid hierarchy to realize a $10\times$~speed-up over AMG for
heterogeneous diffusion
problems~\cite{SPMacLachlan_JMTang_CVuik_2008a}.  In addition, the
communication patterns in a parallel BoxMG solve are fixed and
predictable throughout the multigrid hierarchy.  Here we explore using
this information to improve the parallel scalability and performance
of elliptic solves for problems with structure.

The meshing strategy used in the discretization of a PDE has a
significant impact on the amount of structure that can be exploited by
the solver. For example, single-block locally structured grids can be
body fitted to capture smooth non-planar
geometries~\cite{Delzanno2008}, and can use embedded boundary
discretization techniques to add additional flexibility to the
representation of object boundaries. Robust structured variational
methods, such as BoxMG, are directly applicable to these cases.  In
contrast, fully unstructured grids can capture very complex
geometries, including non-smooth features over a range of scales, but
demand general algebraic multilevel solvers, such as AMG or smoothed
aggregation AMG.  The needs of many applications lie between these
extremes, and a variety of adaptive or multi-mesh strategies have been
developed to preserve structure and enable its use in the
discretization and solver.  These approaches generally lead to
specialized hybrid solvers, favoring structured techniques at higher
levels of refinement and algebraic techniques below a suitably chosen
coarse level~\cite{sundar_parallel_2012}.
While these hybrid solvers present a variety of
challenges, a single-block logically structured solver is a
critical component of their design and performance.

\begin{figure}[ht!]
  \centering \includegraphics[width=.9\textwidth]{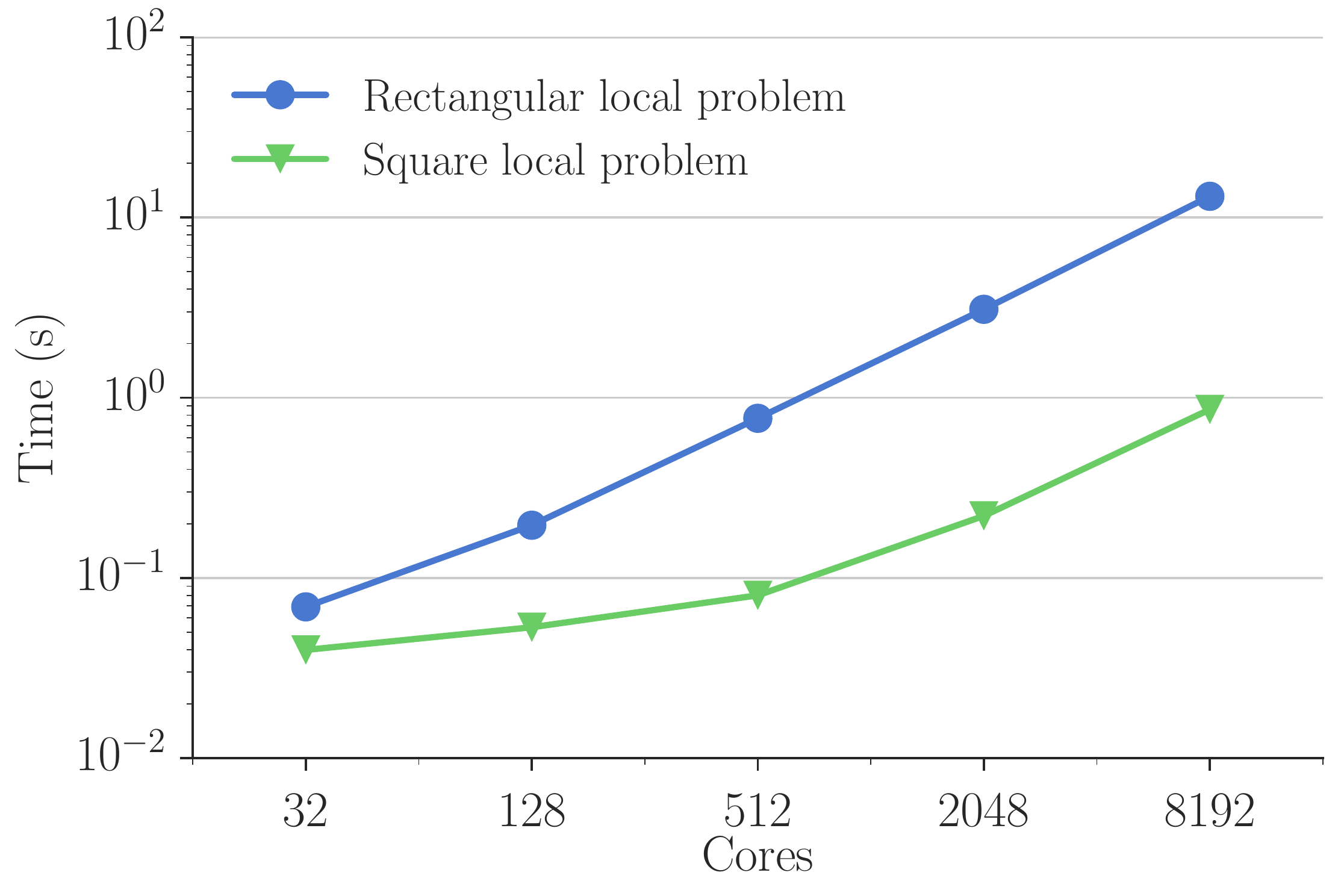}
  \caption{Weak scaling on Blue Waters for a local problem size of
    $568 \times 71$ (rectangular) and $200 \times 200$ (square)
    using V(2,1)-cycles with a gather-to-one serial BoxMG coarse-grid
    solver.}\label{fig:bmgcg}
\end{figure}

A common approach to parallel multigrid solvers for PDEs is to
partition the spatial domain across available processor cores.
However, on coarser levels, the local problem size decreases and the
communication cost begins to impact parallel performance.  A natural
approach to alleviate this problem is to gather the coarsest problem
to a single processor (or redundantly to all the processors), and to
solve it with a serial multigrid cycle.  This approach was first
motivated by a performance model of early distributed memory
clusters~\cite{gropp-1992} where core counts were quite small, and it
was used in the initial version of parallel BoxMG.  Unfortunately, as the
weak scaling study in Figure~\ref{fig:bmgcg} shows, this approach
scales linearly with the number of cores, and hence, is not practical
for modern systems. A modification of this approach that gathers the
coarse problem to a multi-core node~\cite{nakajima-2014}, and then
leverages OpenMP threading on that node, improves the performance but
does not address the scaling. This challenge of reducing communication
costs at coarser levels is even more acute in AMG methods, and this
led to exploration of agglomeration to redundant coarse problems at
higher levels~\cite{Baker2012}, as well as redistribution of smaller
portions of the coarse problem to enhance
scalability~\cite{gahvari-2013}, leading to approximately 2$\times$
speedup in the solve phase.

An alternative to this two-level gather-to-one approach is a more
conservative redistribution strategy that can be applied recursively
as the problem is coarsened.  In single-block structured solvers, the
decision to redistribute data is driven by balancing communication
costs in relaxation with diminishing local work.  This approach was
first considered in a structured setting~\cite{womble_1990}, and used
a heuristic to guide recursive application of nearest neighbor
agglomeration.  Later, in the Los Alamos AMG (LAMG) solver, a
heuristic was developed to guide the reduction of the number of active
cores at each level by a power of two~\cite{Joubert2006}.
In this paper, an optimized redistribution algorithm is proposed for
robust structured multigrid methods that balances the computation and
communication costs at each level.  The structured setting enables the
enumeration of possible coarse-grid configurations, and a performance
model is developed to support optimization of the coarsening path that
is selected through these configurations.  The utility of this
approach is demonstrated through scaling studies extending beyond 100K
cores on two modern supercomputers.

The remainder of this paper is organized as follows.
Section~\ref{sec:multigrid} highlights relevant features of robust
variational multigrid methods, examines the domain decomposition
implementation in BoxMG, and proposes a redistribution algorithm that
can be applied recursively.  The performance model for this parallel
algorithm with redistribution is developed in
Section~\ref{sec:perf_model}, and the optimization algorithm is
presented in Section~\ref{sec:alg}.  Scaling studies are presented in
Section~\ref{sec:experiments} demonstrating the efficacy of the
proposed method, and Section~\ref{sec:conclusions} gives conclusions.

\label{sec:intro}

\section{Robust Variational Multigrid on Structured Grids}\label{sec:multigrid}
In the case of a symmetric, positive definite matrix problem, a
Galerkin (variational) coarse-grid operator is effective
in defining the coarse-level problems because it minimizes the error
in the range of interpolation~\cite{Brand1984}. This variational
operator is formed as the triple matrix product of the restriction
(transpose of interpolation), the fine-grid operator, and the
interpolation, making it well suited for black box and algebraic
multilevel solver algorithms. In the case of structured grids, the
complexity of the coarse-grid operators is bounded, and the stencil
pattern can be fixed \textit{a priori}.

However, in this approach, the interpolation must be sufficiently accurate to
ensure the variational coarse-grid operator satisfies the approximation
property~\cite{Trottenberg:2000}. For example, in two-dimensional
diffusion problems with discontinuous coefficients, bilinear
interpolation is not accurate across discontinuities
because the gradient of the solution is not continuous. Thus, a key
element in robust multigrid methods is \textit{operator-induced}
interpolation~\cite{Briggs:2000,Trottenberg:2000}, which uses the
matrix problem to construct intergrid transfer operators.  In the case
of structured grid problems, operator-induced interpolation is
naturally motivated by noting the normal component of the flux is
continuous~\cite{Dendy1982}, and its impact on the properties of the
variational coarse-grid operator is understood through its connection
to homogenization~\cite{Moulton1998,MacLachlan2006}. This approach
has natural extensions to non-symmetric problems as
well~\cite{Dendy1983,DeZeeuw1990}.

Robust methods also require careful consideration of the smoothing
operator.  Although, Gauss-Seidel is effective for many problems,
anisotropy often demands alternating line smoothing or plane
smoothing~\cite{Briggs:2000,Trottenberg:2000}.  With coarse-grid
operators and interpolation defined, a standard multigrid cycling is
used, for example a V-cycle as in Algorithm~\ref{alg:vcycle}.
\begin{algorithm2e}[ht!]
  \caption{Multilevel V-cycle}\label{alg:vcycle}
  \DontPrintSemicolon	\KwIn{    \begin{tabular}[t]{l l}
       $x_{L-1}$ & fine-grid initial guess \\
       $b_{L-1}$ & fine-grid right-hand side\\
       $A_0,\ldots,A_{L-1}$&\\
       $P_1,\ldots,P_{L-1}$&\\
    \end{tabular}
  }
  \BlankLine	\KwOut{    \begin{tabular}[t]{l l}
       $x_{L-1}$ & fine-grid iterative solution
    \end{tabular}
  }
  \BlankLine  \For{$l = L-1,\ldots,1$}{    \texttt{relax}$(A_l, x_l, b_l, \nu_1)$\tcc*{relax $\nu_1$ times}
    $r_l = b_l - A_l x_l$\tcc*{compute residual}
    $b_{l-1} = P_l^T r_l$\tcc*{restrict residual}
  }
  $x_0 = \ $\texttt{solve}$(A_0, b_0)$\tcc*{coarse-grid direct solve}
  \For{$l = 1,\ldots,L-1$}{    $x_l = x_l + P_l x_{l-1}$\tcc*{interpolate and correct}
    \texttt{relax}$(A_l, x_l, b_l, \nu_2)$\tcc*{relax $\nu_2$ times}
  }
\end{algorithm2e}

In this paper, interpolation and coarse-grid operators are constructed
using the implementation in
BoxMG~\cite{alcouffe1981,Dendy1982,Moulton1998} with standard
coarsening by a factor of 2,
however the approach applies to any structured method.  BoxMG is used
because its operator-induced interpolation is designed to handle
discontinuous coefficients, and additional heuristics ensure
optimal performance for Dirichlet, Neumann and Robin boundary
conditions on logically structured grids of any dimension (i.e.,
it is not restricted to grids $2^k+1$).  In addition,
only the fine-grid problem is specified; coarse-grid operators are
constructed through the variational (Galerkin) product.
The package is also released as open source in the Cedar
Project~\cite{cedar}.

\subsection{Domain Decomposition Parallel Implementation}

\begin{figure}
  \centering
  \includegraphics[width=0.5\textwidth]{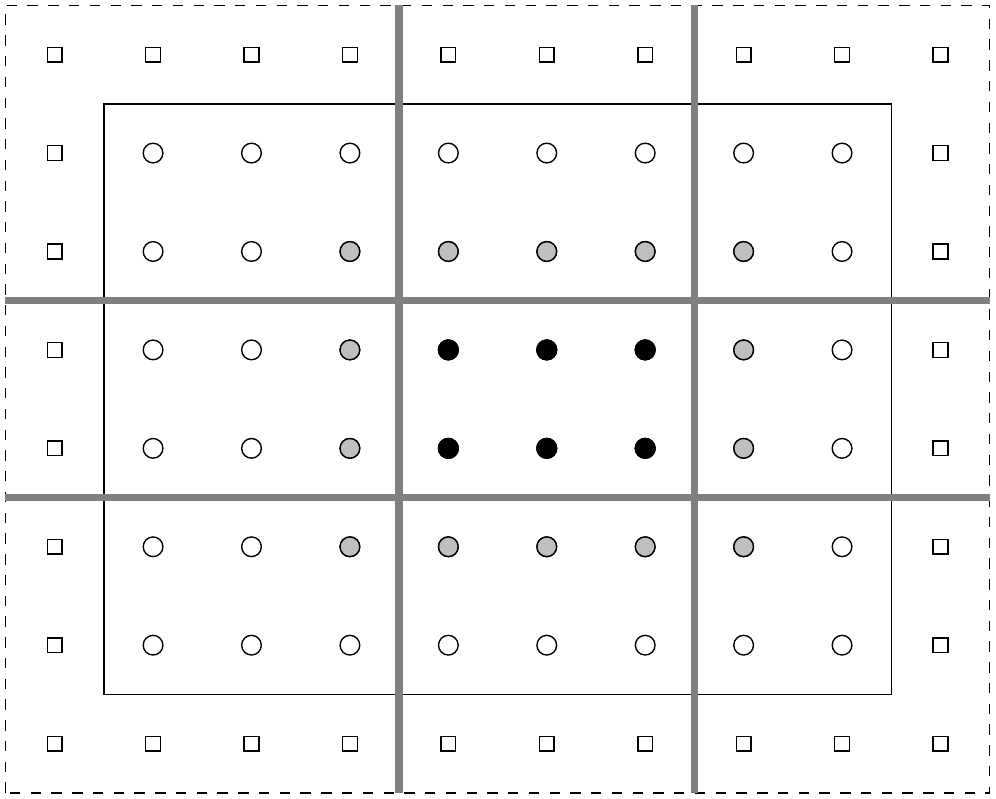}
  \caption{Domain partition.  Circles represent degrees of freedom, with \tikzcircle[fill=black]{2pt} denoting points on process $k$, \tikzcircle[fill=gray!50]{2pt} representing \textit{halo} points needed for communication from other processors, and \tikzcircle[fill=white]{2pt} points on other processors.}\label{fig:procgrid}
\end{figure}

The distributed memory parallel implementation of BoxMG divides the
problem domain among available processors.  The processors are
arranged in a structured grid and points in the fine grid problem are
divided among processors in each dimension~---~see
Figure~\ref{fig:procgrid}.  Since the computation is structured,
inter-process communication occurs through nearest neighbor halo
updates with a halo width of 1.  An important feature of BoxMG is
bounded complexity in the coarse-grid operator.  By exploiting the
structure of the problem, BoxMG produces coarse-grid problems with a
fixed structure.  This results in known communication patterns and
guaranteed data locality.

Examining the distributed memory parallel implementation of BoxMG in
the context of parallel scalability, many of the operations are
stencil based computations with halo updates.  Since the operations
are relatively local, they are not expected to significantly limit
parallel scalability.
For this paper,
the particular focus is in parallel decisions at coarse grids.
To balance useful local work with communication costs,
agglomeration methods are used to redistribute coarse problems.
A straightforward approach is agglomeration to one task.  This task is
then mapped to either a single processor or to all processors
redundantly.  This method of agglomeration was used in the initial distributed
memory implementation of BoxMG\@; while effective for low processor counts
it suffers at scale since the global coarse problem grows linearly with the
number of processors.  As seen in Figure~\ref{fig:bmgcg}, this is especially true with rectangular local
problems
for which the coarsest local problem size is larger, and hence, the linear
growth is faster and overall parallel scaling is lower.
In addition to
agglomerating to one task, the coarse problem can be redistributed by
agglomerating to $n$ tasks, where $n$ is less than the number of
processors.  The problem then becomes choosing a desirable value of
$n$ which is dependent on the problem distribution on the processor
grid.  This agglomeration can be applied recursively to gradually
reduce the size of the processor grid as the size of the global coarse-grid problem is reduced.

\subsection{Redistribution}

To extend parallel coarsening in a structured setting, the algorithm
introduced in this paper
aims to redistribute the coarse-grid problem to an incrementally
smaller subset of processors.  Parallel coarsening continues until a
parameterized minimum local problem size is reached.  At this point,
a set of possible redistributions is enumerated.  These redistributions are
evaluated based on a cost derived through a performance model.  An
optimal redistribution sequence is then selected.  It is important to
note that since a redistributed problem on a given level also limits
parallel coarsening, the selection algorithm is designed to
be applied recursively to obtain the highest efficiency possible
for the multigrid cycle.  Section \ref{ssec:pgrid-enum} discusses
the redistribution of a grid of processors to coarser processor grids.

To redistribute the coarse-grid problem on a smaller number of
processors, the processor grid is agglomerated into processor blocks.
This agglomeration is performed by dimension to maintain a
distributed tensor-product grid structure.  Each processor block is
then mapped to one task in the redistributed solver.

\begin{figure}[ht!]
  \centering
  \includegraphics[width=.6\textwidth]{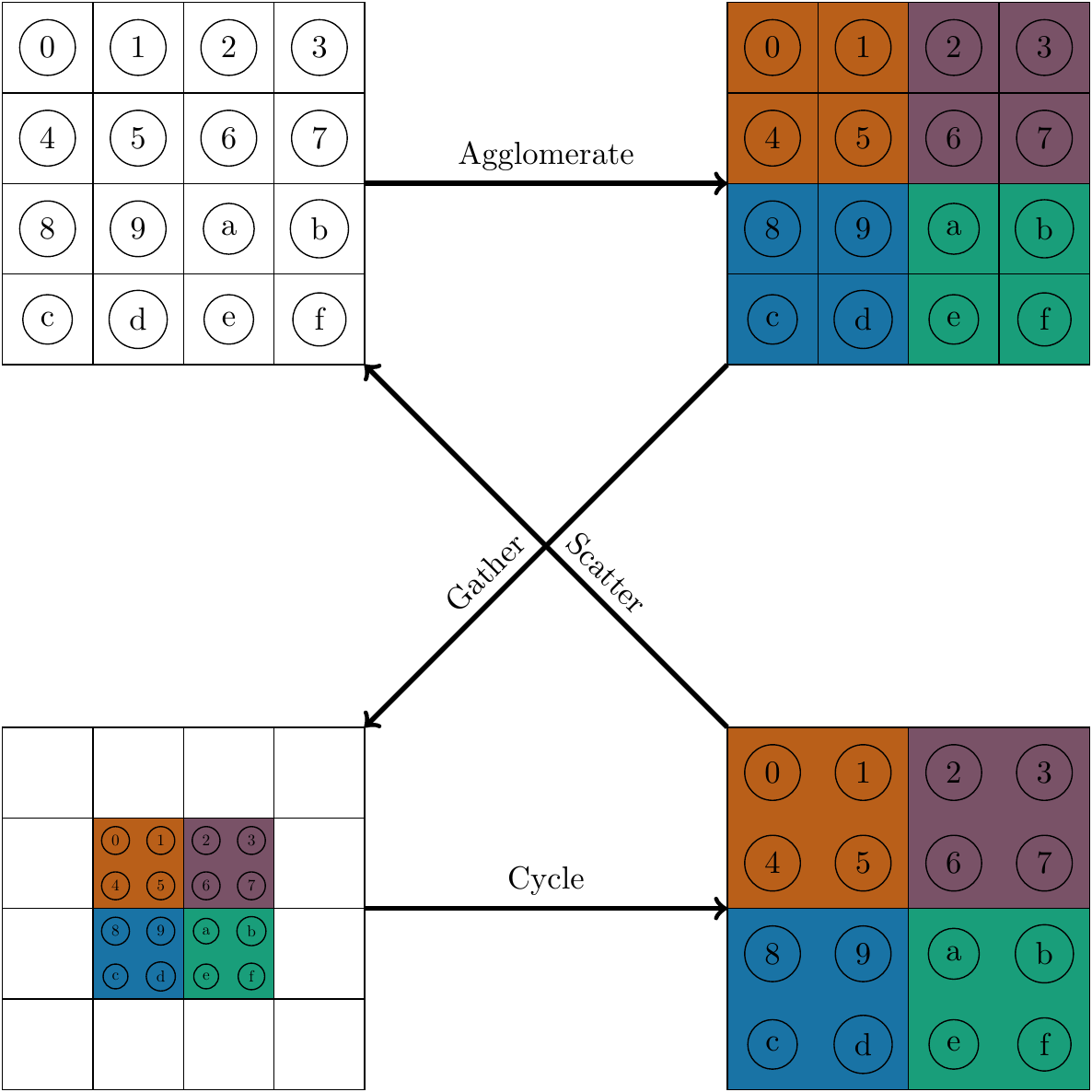}\\
  \caption{A redistribution of a $ 4 \times 4 $ processor
    grid to $2 \times 2$ processor blocks.  The boxes represent the
    processing elements and the circles represent their respective
    coarse-grid local problems.}\label{fig:ex-gs}
\end{figure}

\begin{figure}[ht!]
  \centering
  \includegraphics[width=.6\textwidth]{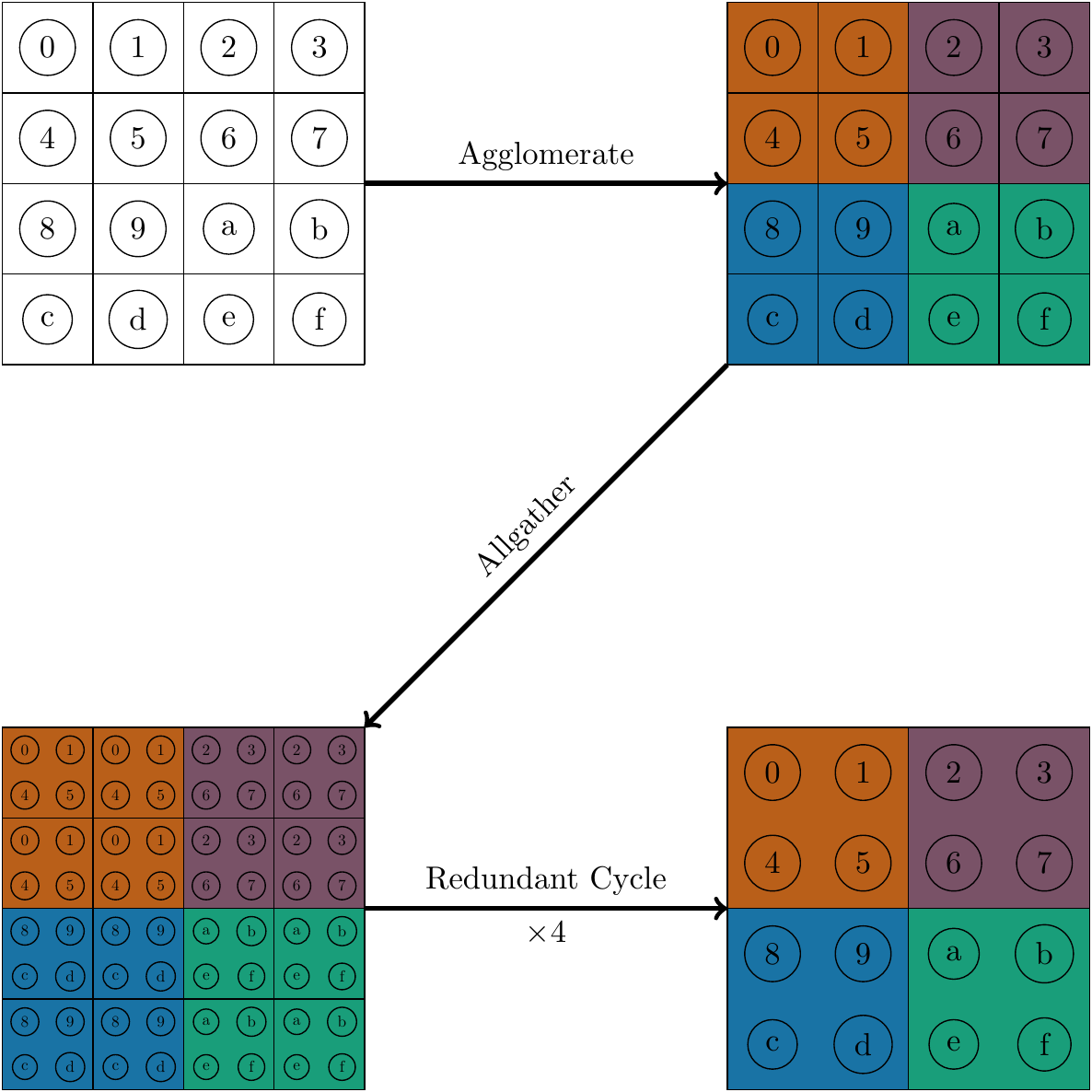}
  \caption{A redundant redistribution of a $ 4 \times 4 $ processor
    grid to $2 \times 2$ processor blocks.  The boxes represent the
    processing elements and the circles represent their respective
    coarse-grid local problems.}\label{fig:ex-allgather}
\end{figure}

To map the
coarse tasks to processors, two approaches are considered.
The first
approach uses one processor from each processor block.
This is visualized in Figure~\ref{fig:ex-gs} where the following steps are taken:
\begin{enumerate}
  \item {\bf Agglomerate:} processors grouped into blocks to define coarse tasks
  \item {\bf Gather:} processors within each block perform gather on coarse problem
  \item {\bf Cycle:} cycling continues with redistributed problem
  \item {\bf Scatter:} iterative solution scattered after redistributed cycle completes
\end{enumerate}
The second approach employs redundancy by
mapping each processor in the processor block to the coarse task.
This is visualized in Figure~\ref{fig:ex-allgather} with the following steps
\begin{enumerate}
  \item {\bf Agglomerate:} processors grouped into blocks to define coarse tasks
  \item {\bf Allgather:} processors within each block perform allgather on coarse problem
  \item {\bf Cycle:} cycling continues redundantly with redistributed problem
\end{enumerate}
While the second approach avoids an additional communication phase at
the end of each cycle and adds an opportunity for resilience through
redundant cycling, the first approach avoids the increased network
usage involved in simultaneous coarse cycles.
Algorithm~\ref{alg:redist} supplements Algorithm~\ref{alg:vcycle} with
steps needed for redistribution.  The algorithm is annotated with
parallel communication required for each step.

\begin{algorithm2e}[ht!]
  \caption{Multilevel V-cycle with Redistribution}\label{alg:redist}
  \DontPrintSemicolon	\KwIn{    \begin{tabular}[t]{l l}
       $x_{L-1}$ & fine-grid initial guess \\
       $b_{L-1}$ & fine-grid right-hand side\\
       $A_0,\ldots,A_{L-1}$&\\
       $P_1,\ldots,P_{L-1}$&\\
       $p_0,\ldots,p_{L-1}$& number of processors on each level\\
    \end{tabular}
  }
  \BlankLine	\KwOut{    \begin{tabular}[t]{l l}
       $x_{L-1}$ & fine-grid iterative solution
    \end{tabular}
  }
  \BlankLine  \For{$l = L-1,\ldots,1$}{      \texttt{relax}$(A_l, x_l, b_l, \nu_1)$\tcc*{halo exchange}
      $r_l = b_l - A_l x_l$\tcc*{halo exchange}
      $b_{l-1} = P_l^T r_l$\;
      \If{$p_l > p_{l-1}$}{        \texttt{gather\_rhs}$(b_{l-1}, p_l, p_{l-1})$\tcc*{local gather}
      }
  }
  $x_0 = \texttt{solve}{(A_0, b_0)}$\;
      \For{$l = 1,\ldots,L-1$}{        \If{$p_l > p_{l-1}$}{          \texttt{scatter\_sol}$(x_{l-1}, p_l, p_{l-1})$\tcc*{local scatter}
         }
      $ x_l = x_l + P_l x_{l-1}$\tcc*{halo exchange}
      \texttt{relax}$(A_l, x_l, b_l, \nu_2)$\tcc*{halo exchange}
    }
\end{algorithm2e}

\section{Performance Model}\label{sec:perf_model}
In this section, a performance model is introduced for the BoxMG V-cycle (see
Algorithm~\ref{alg:redist}).  The model helps identify the parallel performance
limitations of the V-cycle, particularly at coarse-levels in the multigrid
hierarchy, and also provides a cost metric that is used to guide the
coarse-level redistribution algorithm introduced in Section~\ref{sec:alg}.

A key kernel in the V-cycle is that of matrix-vector multiplication.  Since the
stencil-based computations for this operation are relatively uniform, the cost
of parallel communication in  matrix-vector multiplication is accurately modeled with a
\textit{postal} model~\cite{postal}, leading to a
total cost of
\begin{equation}
  T = \underbrace{n_f \cdot \gamma}_{\textnormal{computation}} +
      \underbrace{\alpha + m \cdot \beta}_{\textnormal{communication}},
\end{equation}
with $n_f$ denoting the number of floating point operations, $\gamma$ a measure
of the computation rate or inverse \textit{effective} FLOP rate, $\alpha$ the interprocessor latency, $1/\beta$ the network bandwidth, and
$m$ the number of bytes in an MPI message.  The value $\gamma$ is determined by
measuring the computation time of a local stencil-based matrix-vector product,
while $\alpha$ and $\beta$ are determined through standard machine benchmarks
such as \texttt{mpptest}\footnote{\url{http://www.mcs.anl.gov/research/projects/mpi/mpptest/}}.  As an example, the parameters derived for a 9-point 2D stencil on Blue Waters, a Cray XE6 machine at the National Center for Supercomputing Applications\footnote{\url{https://bluewaters.ncsa.illinois.edu/blue-waters}}, are listed in Table~\ref{tab:bwmodel}.
A more accurate model for communication may be used,
particularly for multiple communicating cores with large message sizes~\cite{2016_GrOlSa_pingpong}
or to account for network contention~\cite{bienzPP16}, which may
play a prominent role in communication.
\begin{table}[!ht]
  \centering
  \begin{tabular}{ccc}
    \toprule
    $\alpha$ & $\beta$ & $\gamma$\\
    \midrule
    \SI{0.65}{\micro\second} & \SI{5.65}{\nano\second}/B & \SI{0.44}{\nano\second}/flop\\
    \bottomrule
  \end{tabular}
  \caption{Model parameters on Blue Waters.}\label{tab:bwmodel}
\end{table}

In an $L$-level multigrid V-cycle, Algorithm~\ref{alg:vcycle} is modeled through
\begin{equation}\label{eq:Vmodel}
  T_{\textnormal{V-cycle}} =
  T_{\textnormal{cgsolve}} +
  \sum_{l=1}^{L-1}
  T_{\textnormal{smooth}}^l +
  T_{\textnormal{residual}}^l +
  T_{\textnormal{restrict}}^l +
  T_{\textnormal{interp}}^l +
  T_{\textnormal{agglomerate}}^l.
\end{equation}

In the following expressions, each component of~\eqref{eq:Vmodel}
represents the \textit{actual} implementation and does not necessarily
form a minimum bound on each portion of the computation.  In the model
parameters below, $D$ denotes the number of dimensions in the problem,
$n_d^l$ the number of local grid points in dimension $d$ on level $l$,
and $n_s$ the number of points in the stencil.  Grid quantities
involved in communication and computation are assumed to be 8 byte
double-precision floating-point numbers.  The dimension of the
parallel decomposition is also assumed to match the dimension of the problem.
The communication required for a halo exchange of width $1$ in $D$ dimensions on
level $l$ is modeled as
\begin{equation}
  T^l_{\textnormal{exchange}}(D) = 2 \cdot D \cdot \alpha
  + 2 \cdot \sum_{d=0}^{D-1} n_d^l \cdot 8 \cdot \beta.
\end{equation}

\begin{figure}
  \centering
  \includegraphics[width=\textwidth]{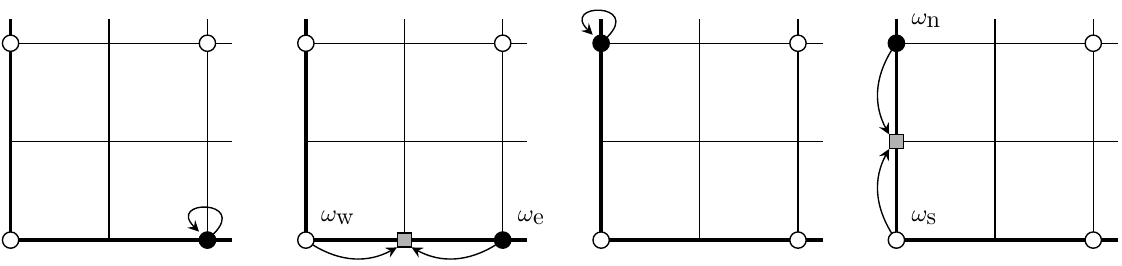}
  \includegraphics[width=.75\textwidth]{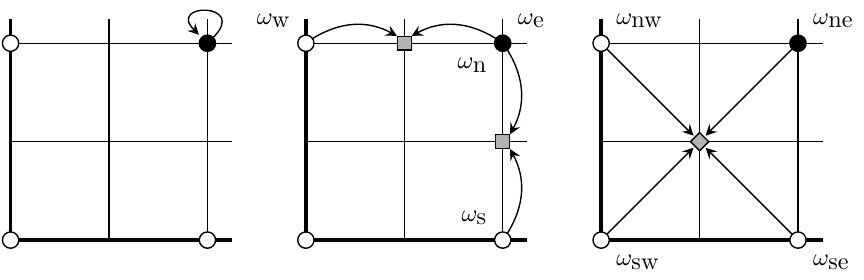}
  \caption{Interpolation computation in 2D.  The top two grids show computation performed on the first coarse line in each dimension.  With the exception of the first, each coarse point performs injection to the corresponding fine point.  Interpolation to the preceding fine point embedded in the coarse-line is then computed using the surrounding coarse points.  The bottom three grids show computation performed for interior coarse points.  For each coarse point, the corresponding fine point is injected.  The preceding coarse points in each dimension are then interpolated.  Finally, the logical cell center preceding the coarse point is interpolated using the surrounding coarse points.
  }\label{fig:interpexample}
\end{figure}

Smoothing, using Gauss-Seidel with $n_c$ colors, results in
\begin{equation}
T_{\textnormal{smooth}}^l = 2 \cdot n_s \cdot \prod_{d=0}^{D-1} n_d^l \cdot (\nu_1 + \nu_2) \cdot \gamma
                          + n_c \cdot (\nu_1 + \nu_2) \cdot T^l_{\textnormal{exchange}}(D).
\end{equation}
Likewise, the residual computation is
\begin{equation}
  T_{\textnormal{residual}}^l = 2 \cdot n_s \cdot \prod_{d=0}^{D-1} n_d^l \cdot \gamma
                  + T^l_{\textnormal{exchange}}(D).
\end{equation}
Since it is unnecessary to communicate halo regions in restriction,
the computation is entirely local, leading to
\begin{equation}
T_{\textnormal{restrict}}^l = 2 \cdot n_s \cdot \prod_{d=0}^{D-1} n_d^l
\cdot \gamma.
\end{equation}
Following~\cite{Dendy2010}, the interpolation and correction computes
\begin{align}
  u^l & \gets u^l + I_{l-1}^l u^{l-1} + r^l/C,\\
\intertext{where $r^l$ is the previously computed residual and $C$ is the
center, diagonal stencil coefficient of the operator.  Interpolation for edges (see Figure~\ref{fig:interpexample}) yields
}
u^l & \gets
\underbrace{u^l +
\omega_\west u^{l-1}_\west +
\omega_\east u^{l-1}_\east +
r^l/C
}_{5 FLOPs}\\
\intertext{for the $x$-direction (and similar for the $y$-direction).  Likewise, the interior stencil (see Figure~\ref{fig:interpexample}) yields}
u^l & \gets
\underbrace{u^l +
\omega_{\southwest} u^{l-1}_{\southwest} +
\omega_{\southeast} u^{l-1}_{\southeast} +
\omega_{\northwest} u^{l-1}_{\northwest} +
\omega_{\northeast} u^{l-1}_{\northeast} +
r^l/C
}_{9 FLOPs}
\end{align}
Here interpolation is given as weights stored at the coarse points.
For example, the interpolation stencil stored at a given coarse point
is
\[
 \begin{bmatrix}
    \omega_{\northwest} & \omega_{\north} &  \omega_{\northeast}\\
    \omega_{\west} &  & \omega_{\east}\\
    \omega_{\southwest} & \omega_{\south} & \omega_{\southeast}
  \end{bmatrix}.
\]
In total (with injection), interpolation in 2D is modeled as
\begin{equation}
T_{\textnormal{interp}}^l = \left( \prod_{d=0}^1 n_d^l + 20 \cdot \prod_{d=0}^1
n_d^{l-1} + 6 \cdot \sum_{d=0}^1 n_d^{l-1} \right) \cdot \gamma + T^l_{\textnormal{exchange}}(2).
\end{equation}
Using a similar derivation, interpolation in 3D is
\begin{equation}
\begin{split}
T_{\textnormal{interp}}^l & = \left( \prod_{d=0}^2 n_d^l + 60 \cdot \prod_{d=0}^2
n_d^{l-1} + 15 \cdot n_0^{l-1} \cdot n_2^{l-1}
+ 6 \cdot n_1^{l-1} \cdot n_2^{l-1} + n_2^{l-1} \right)  \cdot \gamma\\ &+ T^l_{\textnormal{exchange}}(3).
\end{split}
\end{equation}
To agglomerate the coarse problem, the right hand side is gathered
within processor blocks before the redistributed cycle begins and
approximate solution scattered after the redistributed cycle completes.
This time is then given by:
\begin{equation}
  T_{\textnormal{agglomerate}}^l =
  \begin{cases}
  T_{\textnormal{gather\_rhs}}^l + T_{\textnormal{scatter\_sol}}^l & \textnormal{if $p^l > p^{l-1}$}\\
    0 & \textnormal{else},
  \end{cases}
\end{equation}
where $T_{\textnormal{gather\_rhs}}^l$ and $T_{\textnormal{scatter\_sol}}^l$ represent the time to
gather and scatter the right-hand side and solution, as follows.
The number of processors within a processor block and the local
problem size for a processor in a processor block is given as
\[
p_{\textnormal{block}}^l = \prod_{d=0}^{D-1} \left\lceil \frac{p_{d}^{l+1}}{p_{d}^{l}} \right\rceil,\
n_{\textnormal{block}}^l = \prod_{d=0}^{D-1} \left\lceil \frac{N_{d}^l}{p_d^{l}} \right\rceil.
\]
Then the solution an MPI \texttt{allgather} or \texttt{gather/scatter}
depends on whether the redistribution is redundant.
Following~\cite{thakur_optimization_2005},
the cost of these collective operations is given by
\begin{equation}
T_{\textnormal{gather\_rhs}}^l = \lceil \log_2(p_{\textnormal{block}}^l) \rceil \cdot \alpha + n_{\textnormal{block}}^l \cdot \frac{p_{\textnormal{block}}^l - 1}{p_{\textnormal{block}}^l} \cdot 8 \beta
\end{equation}

\begin{equation}
T_{\textnormal{scatter\_sol}}^l =
\begin{cases}
  0 & \textnormal{if redundant}\\
   T_{\textnormal{gather\_rhs}}^l & \textnormal{else}.
\end{cases}
\end{equation}

Finally, the time for a coarse solve is the time to agglomerate to one
processor combined with the cost of a Cholesky direct solve:
\begin{equation}
  T_{\textnormal{cgsolve}} = T_{\textnormal{agglomerate}}^0 + {\left(\prod_{d=0}^{D-1} N_d^0 \right)}^2 \cdot \gamma.
\end{equation}
This assumes the Cholesky factorization has been computed and stored
in the setup phase.

This performance model provides a basic predictive model to begin
exploring the guided redistribution algorithm proposed in this paper.

\section{Optimized Parallel Redistribution Algorithm}\label{sec:alg}

\subsection{Coarse Processor Grid Enumeration}\label{ssec:pgrid-enum}

To enumerate potential redistributions,
the fine-grid tasks described by the processor grid are agglomerated into coarser tasks called
processor blocks.  In agglomerating by dimension, the
processor blocks form a coarser tensor product grid.  The
coarse processor grids are enumerated by beginning with a 1-D grid and by refining greedily
by dimension with respect to the agglomerated local problem size.
Using a refinement factor of $2$, the number of enumerated processor
grids is bounded by $\lceil \log_2{n_p} \rceil$.
\begin{table}[!htb]
  \centering
  \begin{tabular}{cc}
    \toprule
    Redistribution & Agglomerated Local Problem\\
    \midrule
    $1  \times 1$ & $1136 \times 71$\\
    $2  \times 1$ & $568  \times 71$\\
    $4  \times 1$ & $284  \times 71$\\
    $8  \times 1$ & $142  \times 71$\\
    $16 \times 1$ & $71   \times 71$\\
    $16 \times 2$ & $71   \times 36$\\
    $16 \times 4$ & $71   \times 18$\\
    \bottomrule
  \end{tabular}
  \caption{Example redistribution enumeration using a fine grid
    problem of $9088 \times 568$ degrees of freedom with a $16 \times 8$
    processor grid.  The global coarse-grid considered for agglomeration
    contains $1136 \times 71$ degrees of freedom.}
  \label{tbl:enum}
\end{table}
Table~\ref{tbl:enum} illustrates this enumeration strategy using an
initial fine grid problem of $9088 \times 568$ degrees of freedom with
a $16 \times 8$ processor grid.
This refinement
procedure is used to limit the number of potential redistributions thereby
making the global search feasible within the setup phase.

\subsection{Redistribution Search}

\begin{figure}[htbp]
  \centering
    \includegraphics[width=.7\textwidth]{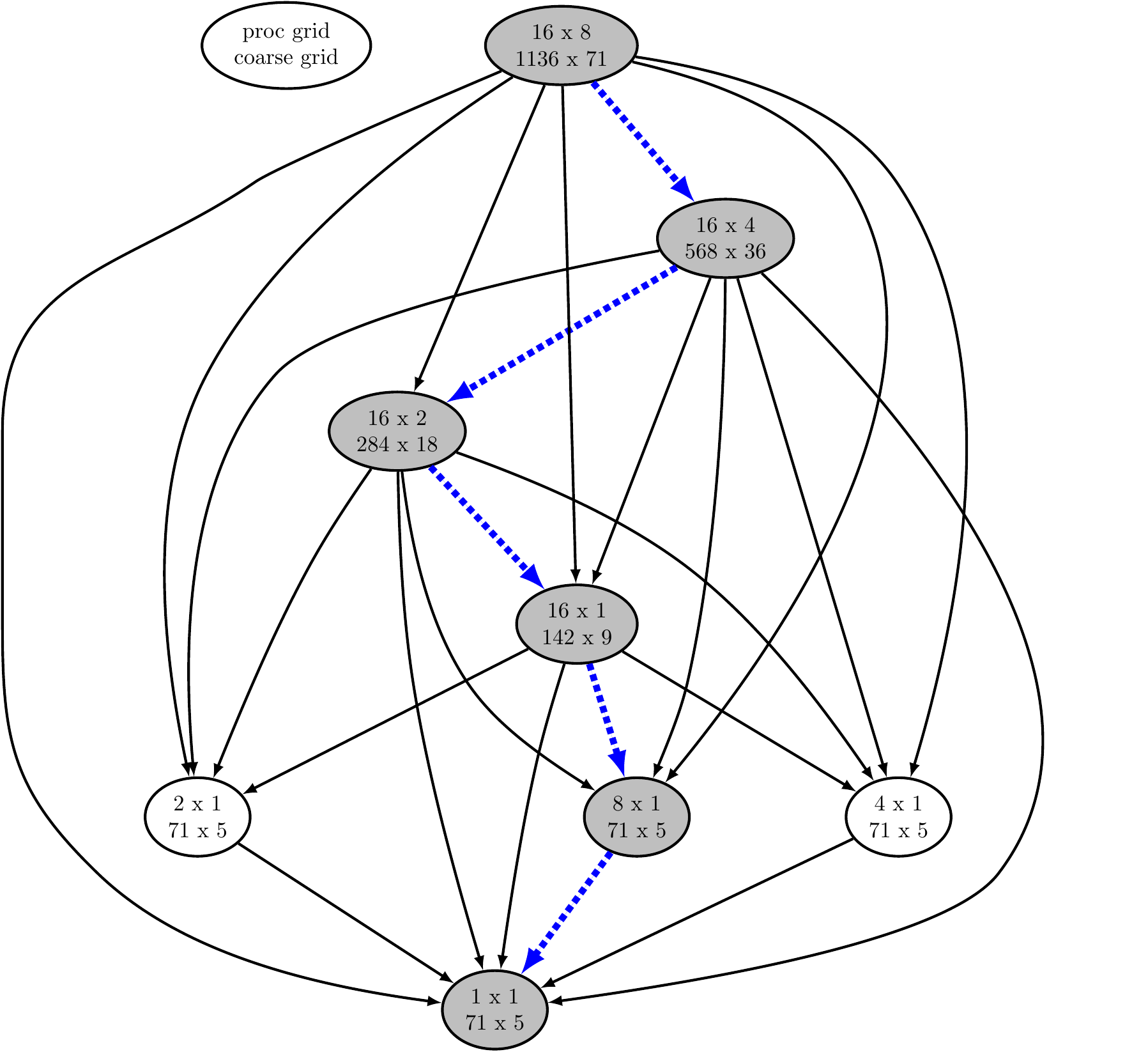}
  \caption{Example redistribution search space: The fine grid global
    problem is $9088 \times 568$ degrees of freedom with a $16 \times 8$ processor grid, yielding a fine grid local problem
    of $568 \times 71$.
    The optimal path through this redistribution space is
    highlighted.}\label{fig:sgraph}
\end{figure}

The recursive enumeration of coarse processor grids generates a search
space of possible redistributions.  To find an optimal
redistribution, a state in the search space is given by the
size of the processor grid and the size of the coarse-grid associated
with a distributed solver.
The search space can be viewed as a directed graph~---~an example is given in
Figure~\ref{fig:sgraph}. The initial state is given by the top-level
distributed solver and the goal state is the state with a $1\times 1$
processor grid (in the case of a 2-D problem).  State transitions have varying costs; the goal is an inexpensive
path from the initial state to the goal state.

To search the state space, a path cost
function is defined as
\begin{equation}
f(s) = g(s) + h(s)
\end{equation}
where $s$ is a vertex or state in the graph in
Figure~\ref{fig:sgraph}, $g$ is the cost to reach $s$ from the initial
state, and $h$ is an estimate of the cost to reach the goal state from
state~$s$.  That is, $g(s)$ represents the time predicted by the
performance model for a solve phase executed down to coarse-grid state
$s$.  For the heuristic function $h(s)$, a weighted combination of the
coarse-grid problem and the processor grid size is used to predict the
cost.
If the performance model is an accurate predictive model, an
inexpensive path from the initial state to the goal state will
identify a redistributed solver with an efficient run-time.  With the
path cost function defined, the space of redistributions is searched
for an optimal path.  This is performed in the MG setup phase to
dictate agglomeration when a coarsening limit is reached.  Using a brute force approach by searching
every path for the best redistribution strategy incurs an
$\mathcal{O}(n_p)$ cost, since the redistribution search space is
constructed recursively.  Letting $h = \log_2(n_p)$, and using the above
coarse processor grid enumeration, results in
\begin{equation}
T(0) = 1, \ T(h) = \sum_{k=0}^{h-1} T(k) + 1.
\end{equation}
Since $2^0 =
T(0) = 1$ and since
\begin{align*}
  T(h+1) &= \sum_{k=0}^h T(k) + 1\\
         &= T(n) + \sum_{k=0}^{h-1} T(k) + 1\\
         &= 2^h + 2^h = 2^{h+1},
\end{align*}
this leads to $T(h) = 2^h = 2^{\log_2(n_p)} = n_p$.

\begin{figure}[hbt!]
  \centering
  \includegraphics[width=.9\textwidth]{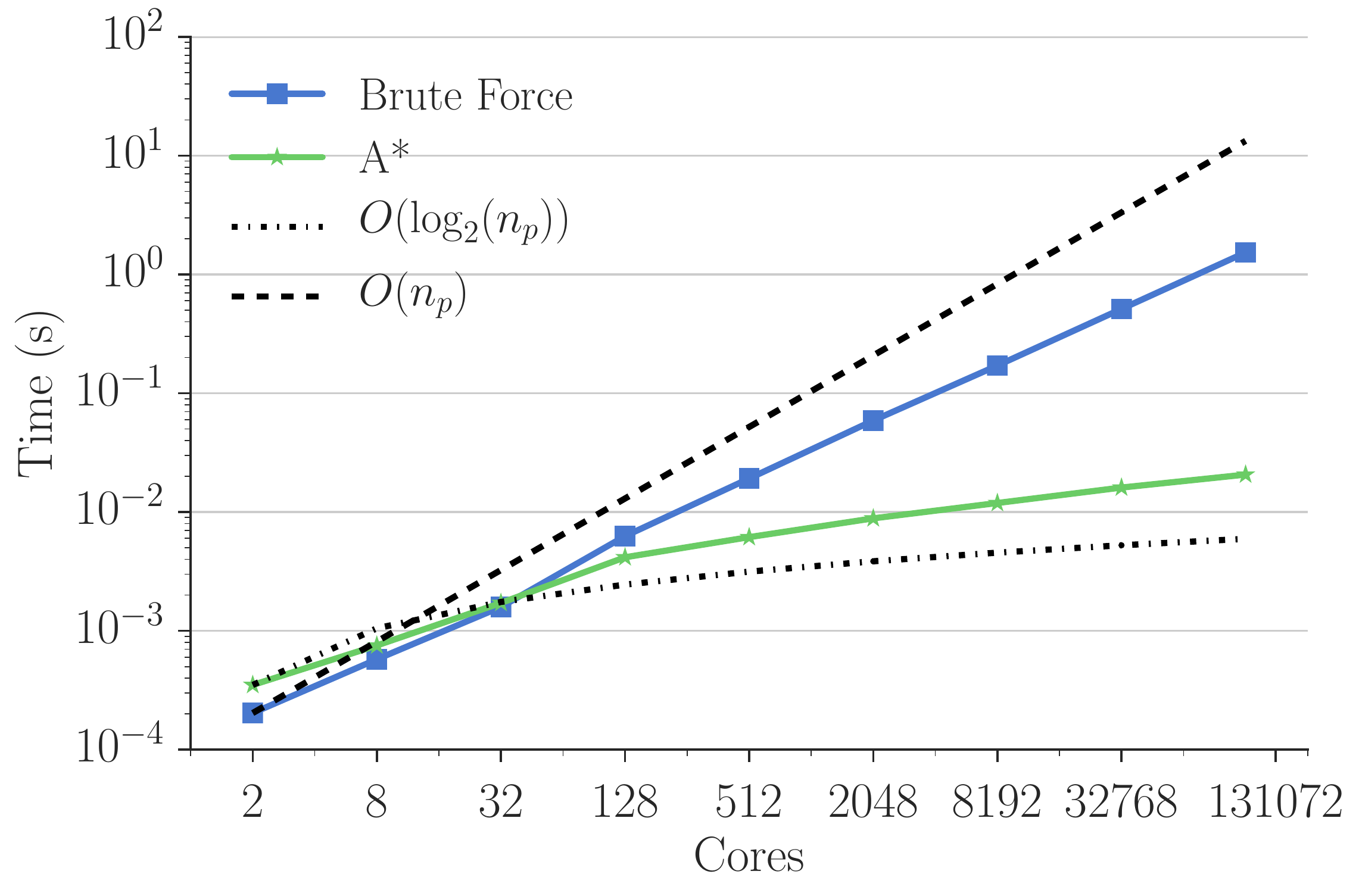}
  \caption{Performance of redistribution search algorithms weak
    scaling with local problem size $568 \times 71$.}\label{fig:stimes}
\end{figure}

To address the $\mathcal{O}(n_p)$ cost, the A*
algorithm is used to determine the optimal path.  While the worst case complexity for A* is
is $\mathcal{O}(n_p)$ for this search, the cost with an optimal heuristic with evaluation cost
$\mathcal{O}(1)$ is the length of the solution path.  The length of this path
in the redistribution search is $\mathcal{O}(\log_2(n_p))$.
Figure~\ref{fig:stimes} shows the A* heuristic is effective
in avoiding the brute force complexity, but does not reach the
optimal complexity.

\section{Experimental results}\label{sec:experiments}

To explore the performance of the proposed redistribution algorithm a
standard 5-point finite volume discretization of the diffusion equation
is used.
Since square problems may lead to a natural or predictable
redistribution path, rectangular local problems are used in order to
fully stress the redistribution algorithm in a weak scaling study.
In particular, the ratio of processors in the processor grid is fixed
at $2\,:\, 1$, and the ratio of unknowns in the fine-grid local
problem is fixed at $8\,:\,1$.  Other processor grid ratios lead to
similar findings.

Note that for an isotropic diffusion problem this grid stretching
results in anisotropy in the discrete problem, causing the convergence
of the solver to deteriorate when using point-wise smoothing.
Consequently, a diffusion problem with compensating
\textit{anisotropy} is defined in order to focus on the parallel
scalability of the coarse-grid redistribution algorithm (rather than
the well-established convergence rate of the multigrid algorithm
itself).  Specifically, the following diffusion problem is used in
these numerical experiments,
\begin{align}
  - \nabla \cdot (D \nabla u)
  &= f(x,y) \quad \text{in} \ \Omega=(0,1) \times (0,1) \, ,
  \label{eq:diffusion} \\
  u &= 0 \quad \text{on} \ \partial \Omega \, ,
  \label{eq:diffusion_bc}
\end{align}
where the diffusion tensor is
$D=\text{diag} [ \frac{1}{r}, r ]$ and $r \approx 16$. This
compensating anisotropy results in optimal convergence of multigrid
V-cycles with point-wise smoothing.

\subsection{Scaling Studies}

Both the weak and strong scalability of multigrid V-cycles that use
the proposed redistribution algorithm are important for applications
on exascale systems.  Here, the discretization of the diffusion
problem given above is used for both weak and strong scaling studies
and the algorithmic components of the BoxMG multigrid library (e.g.,
interpolation, restriction) are used in the redistributed multigrid
V-cycles.

Since, the cost of coarsening (with redistribution) and the cost of
the coarse-grid problem are dominated by parallel communication,
network speed and machine topology play an important role in timings.
To explore this dependency, two different petascale systems are
considered in the scaling tests:
\begin{description}
  \setlength{\itemsep}{6pt}
  \item[Mira\footnotemark]\footnotetext{\url{https://www.alcf.anl.gov/mira}} An IBM Blue Gene/Q system at Argonne National Laboratory.  Mira uses an IBM network which comprises a 5D torus with 49,152 compute nodes using PowerPC A2 processors.
  \item[Blue Waters\footnotemark]\footnotetext{\url{https://bluewaters.ncsa.illinois.edu/blue-waters}} A Cray XE system at the National Center for Supercomputing Applications (NCSA) at the University of Illinois at Urbana-Champaign.  Blue Waters employs a 3D torus using a Cray Gemini interconnect and has 22,640 XE compute nodes each with two AMD Interlagos processors.
\end{description}
In each case, the machine parameters used in the performance model of
Section~\ref{sec:perf_model} are determined using the
b\textsubscript{eff} benchmark~\cite{beff}.

\subsubsection*{Weak Scalability}

In this section, a weak scaling study is conducted to highlight the
scalability of the redistribution algorithm at large core counts.  For
the numerical experiments below, 10 V{(2,1)}-cycles are executed using
two different local problems sizes: $568 \times 71 = 40,328$ and $288
\times 36 = 10,368$.

\begin{figure}[hbt!]
  \centering
  \includegraphics[width=.9\textwidth]{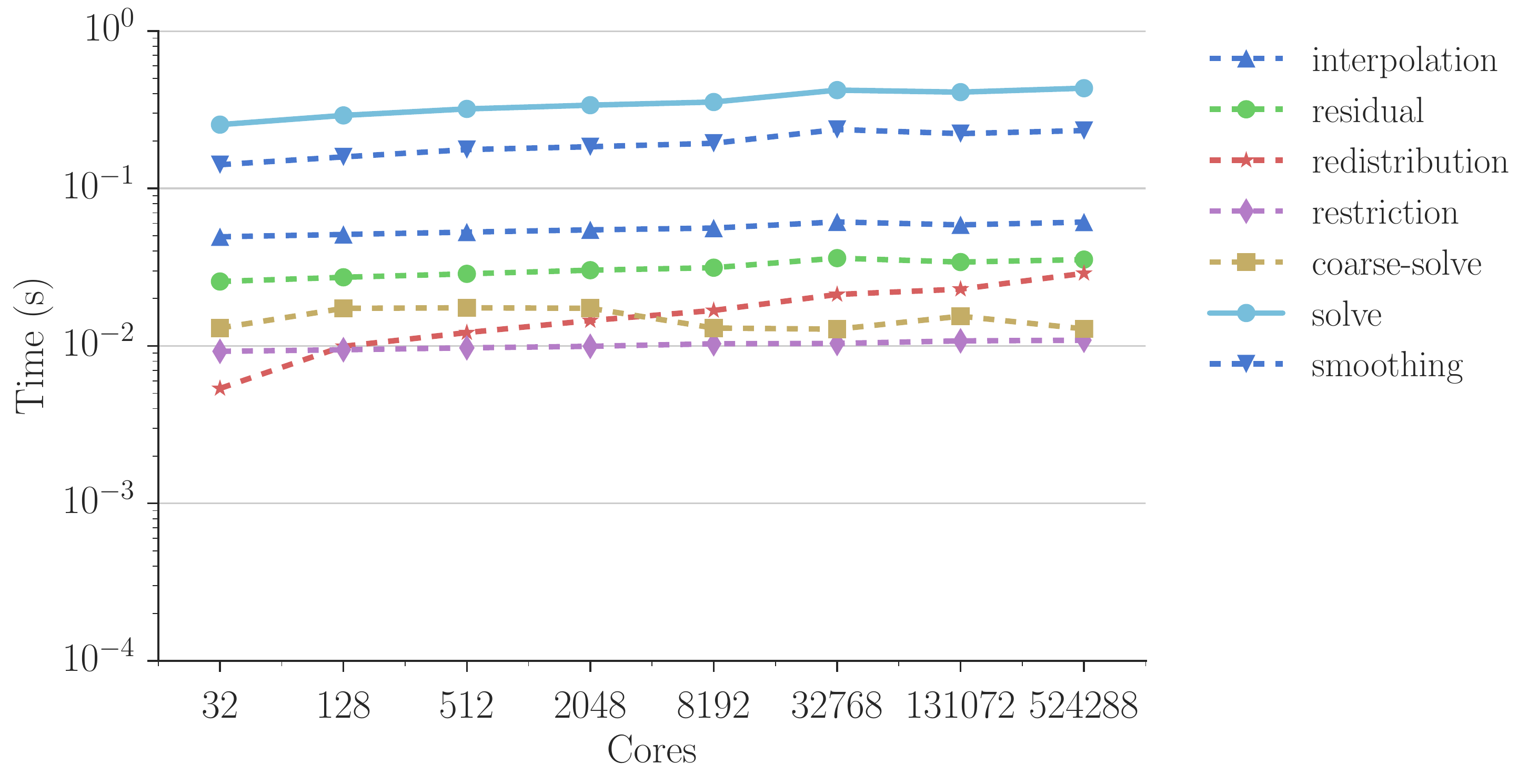}
  \caption{Weak scaling on Mira with local problem size: $568 \times 71$.}\label{fig:mira-40k}
\end{figure}

\begin{figure}[hbt!]
  \centering
  \includegraphics[width=.9\textwidth]{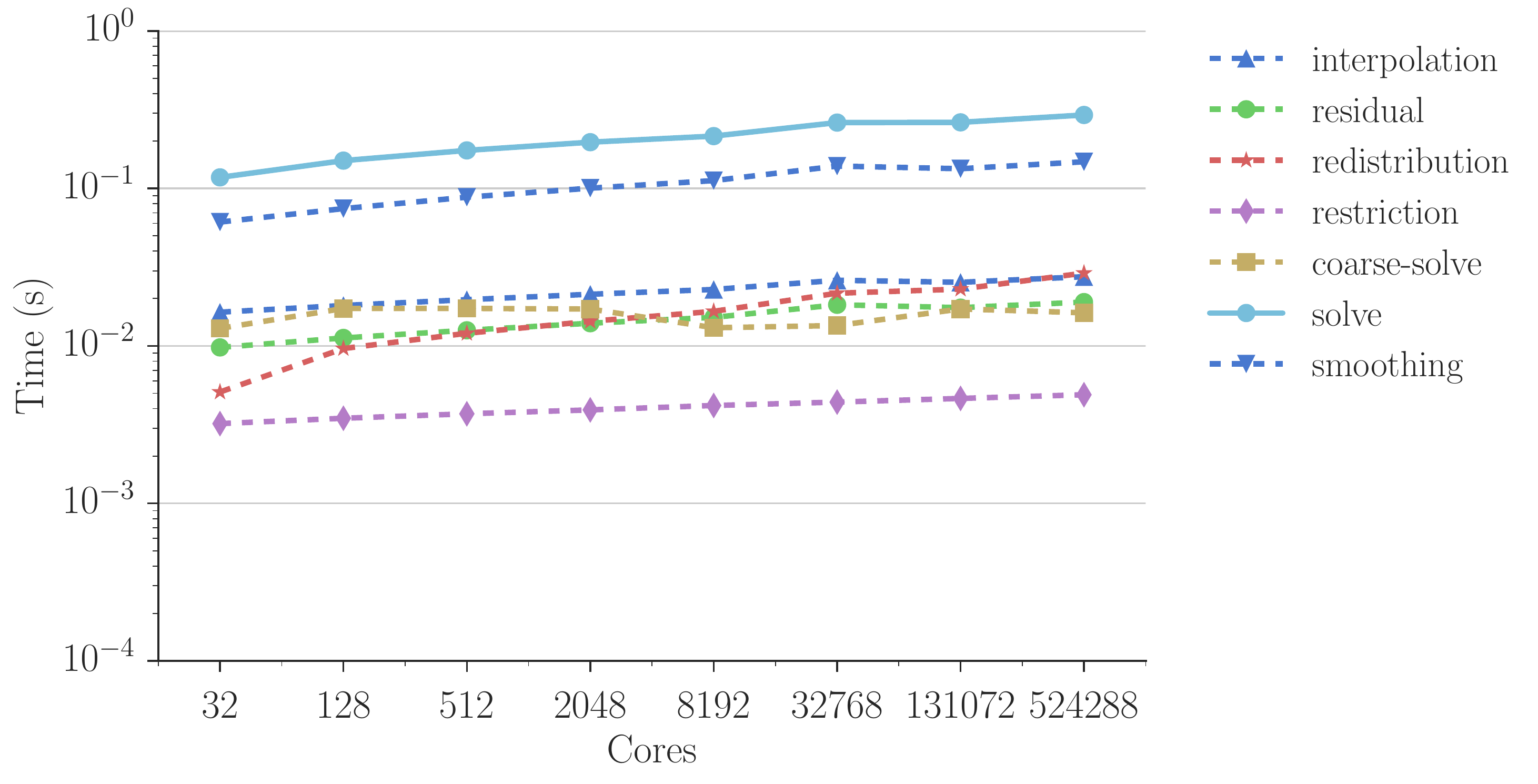}
  \caption{Weak scaling on Mira with local problem size: $288 \times 36$.}\label{fig:mira-10k}
\end{figure}

Figures~\ref{fig:mira-40k}~and~\ref{fig:mira-10k} show the run times on
Mira of various computational kernels in the multigrid solve phase.
The redistribution line shows communication needed to redistribute
coarse problems.  This communication is low in comparison to the
cost of relaxation.  Overall, the algorithm exhibits high parallel scalability in the solve time for both local
problem sizes on Mira.

\begin{figure}[hbt!]
  \centering
  \includegraphics[width=.9\textwidth]{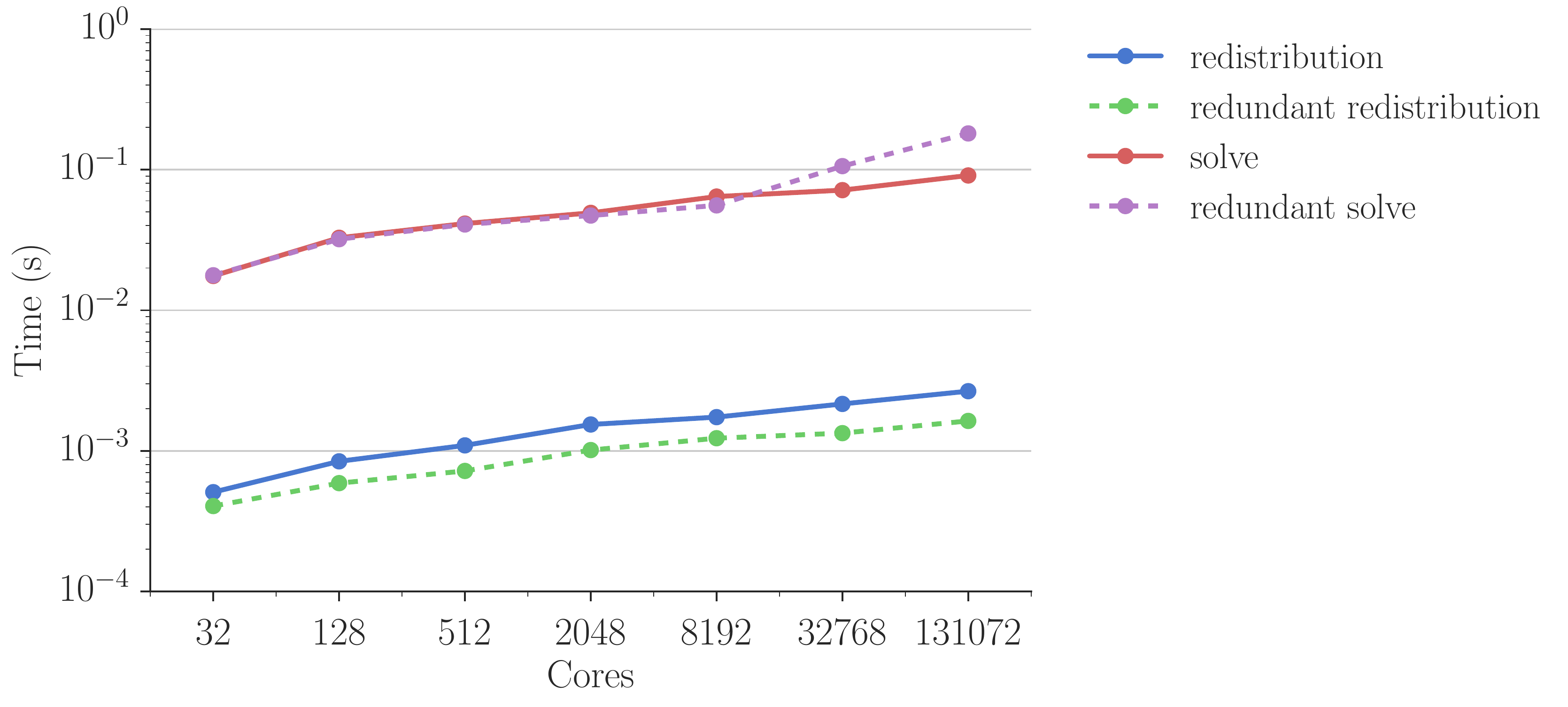}
  \caption{Weak scaling on Blue Waters with local problem size $288
    \times 36$ shows the significant improvement of non-redundant redistribution
    compared to redundant redistribution beyond 8192 cores.}\label{fig:redundant-vs-gs}
\end{figure}

With different network capacities, \textit{redundant} redistribution may not yield
the lowest communication costs.  Indeed, the results in
Figure~\ref{fig:redundant-vs-gs} for the
Blue Waters system show that while redundant redistribution of the data at course levels
is inexpensive, the network contention introduced by redundant cycling
contributes to an increase in communication at high core counts.  This suggests
that triggering redundancy on a per-level basis could lead to reduced costs.
In contrast, non-redundant redistribution (in
Figure~\ref{fig:redundant-vs-gs}) exhibits high scalability~---~thus, it is used for the
following runs on Blue Waters.

\begin{figure}[hbt!]
  \centering
  \includegraphics[width=.9\textwidth]{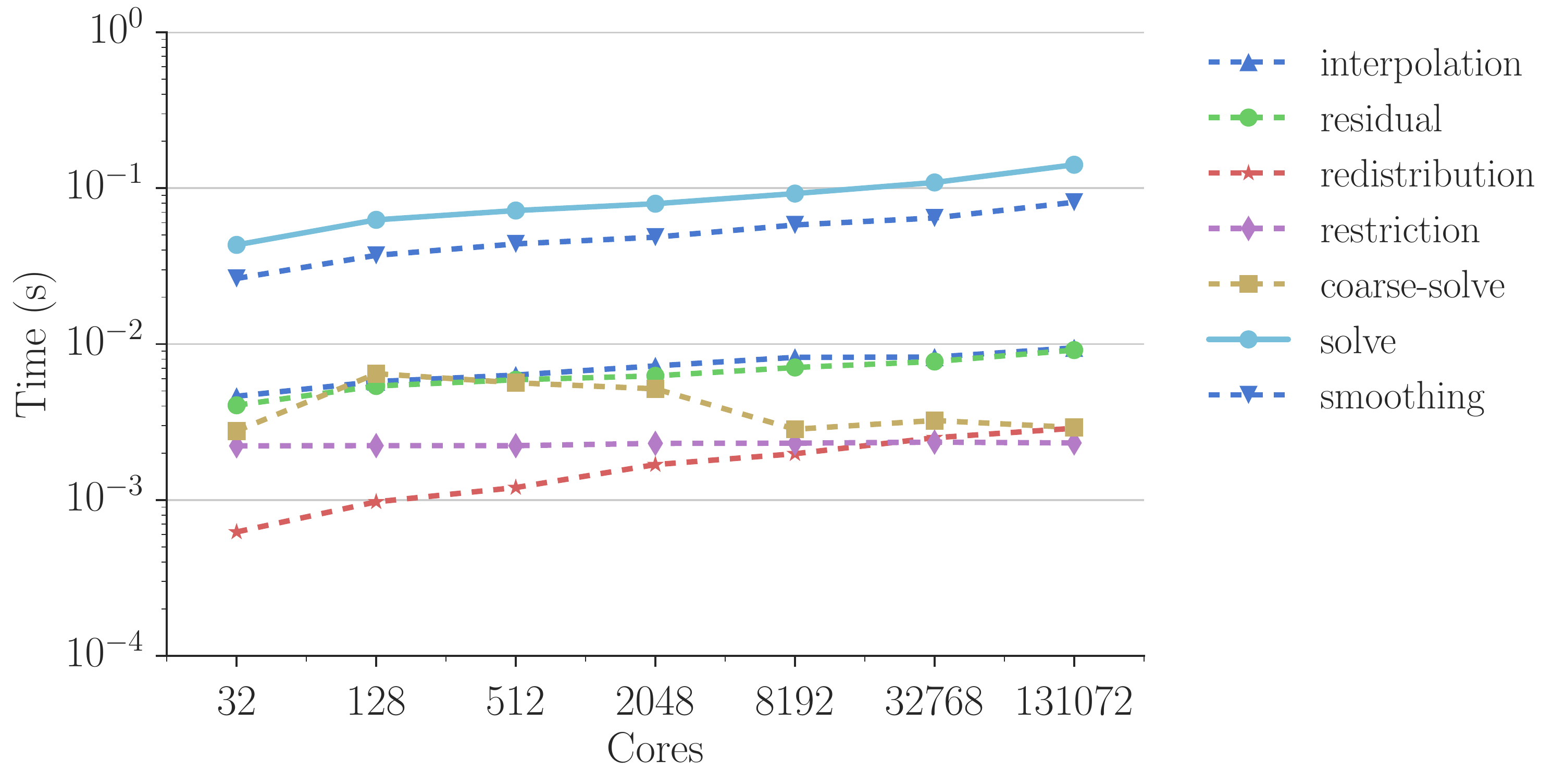}
  \caption{Weak scaling on Blue Waters with local problem size: $568 \times 71$.}\label{fig:bw-40k}
\end{figure}

\begin{figure}[hbt!]
  \centering
  \includegraphics[width=.9\textwidth]{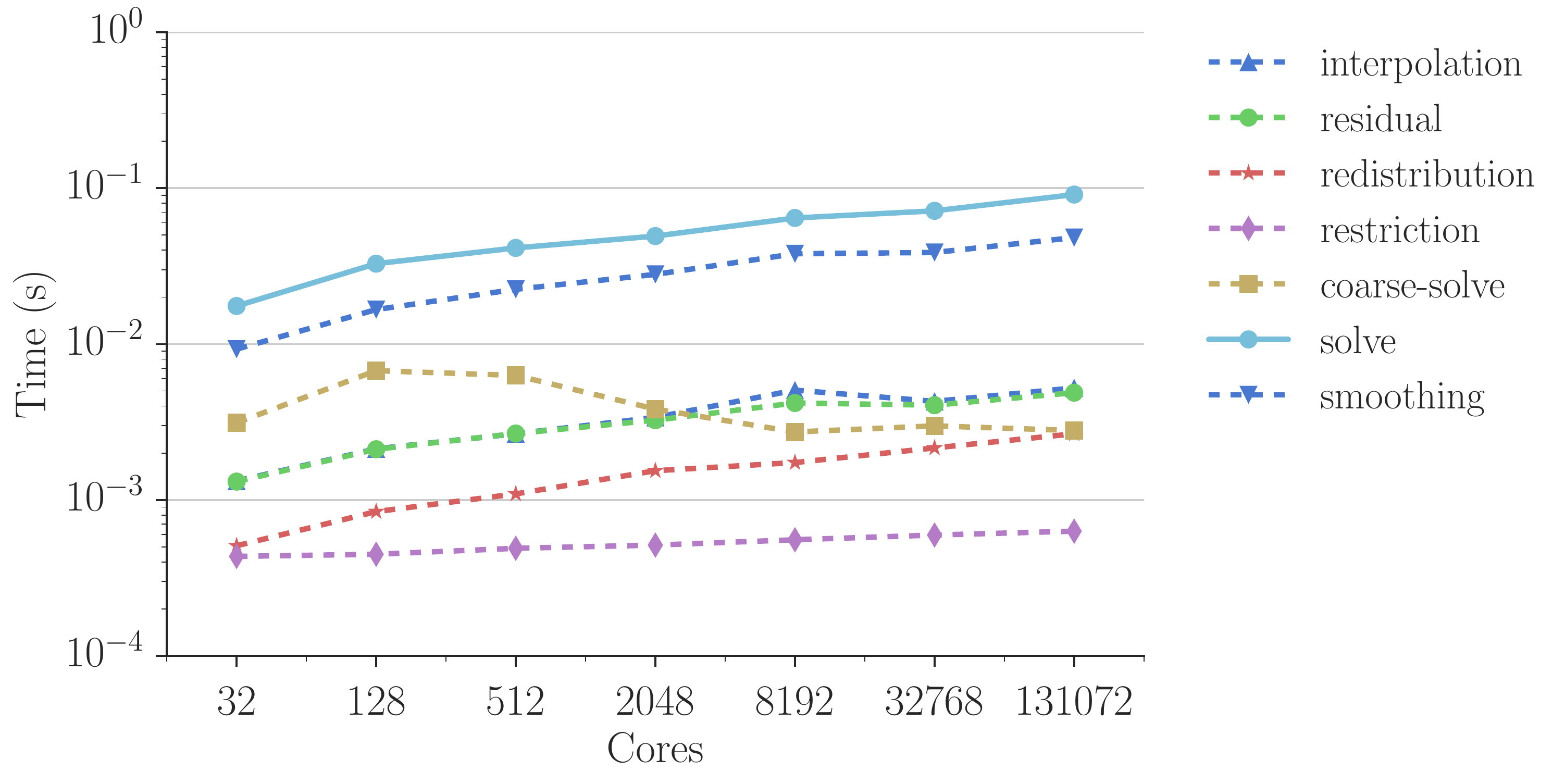}
  \caption{Weak scaling on Blue Waters with local problem size: $288 \times 36$.}\label{fig:bw-10k}
\end{figure}

Figures~\ref{fig:bw-40k}~and~\ref{fig:bw-10k} show run times on
Blue Waters for the multigrid solve phase, highlighting that the
proposed algorithm achieves good parallel scalability here as well.
However, comparing runs on the two machines, it is apparent that the
weak scalability is superior on Mira.  This is attributed to the
network capabilities and scheduling differences of the two machines.
Jobs on Mira receive a dedicated, full torus partition of the machine,
which often reduces contention resulting from neighboring jobs on the
machine, since partitions receive a full torus network.  Moreover, the
lower dimensional 3D torus on Blue Waters also contributes to an
increase in contention within the running job.

\begin{figure}
  \centering
  \includegraphics[width=.9\textwidth]{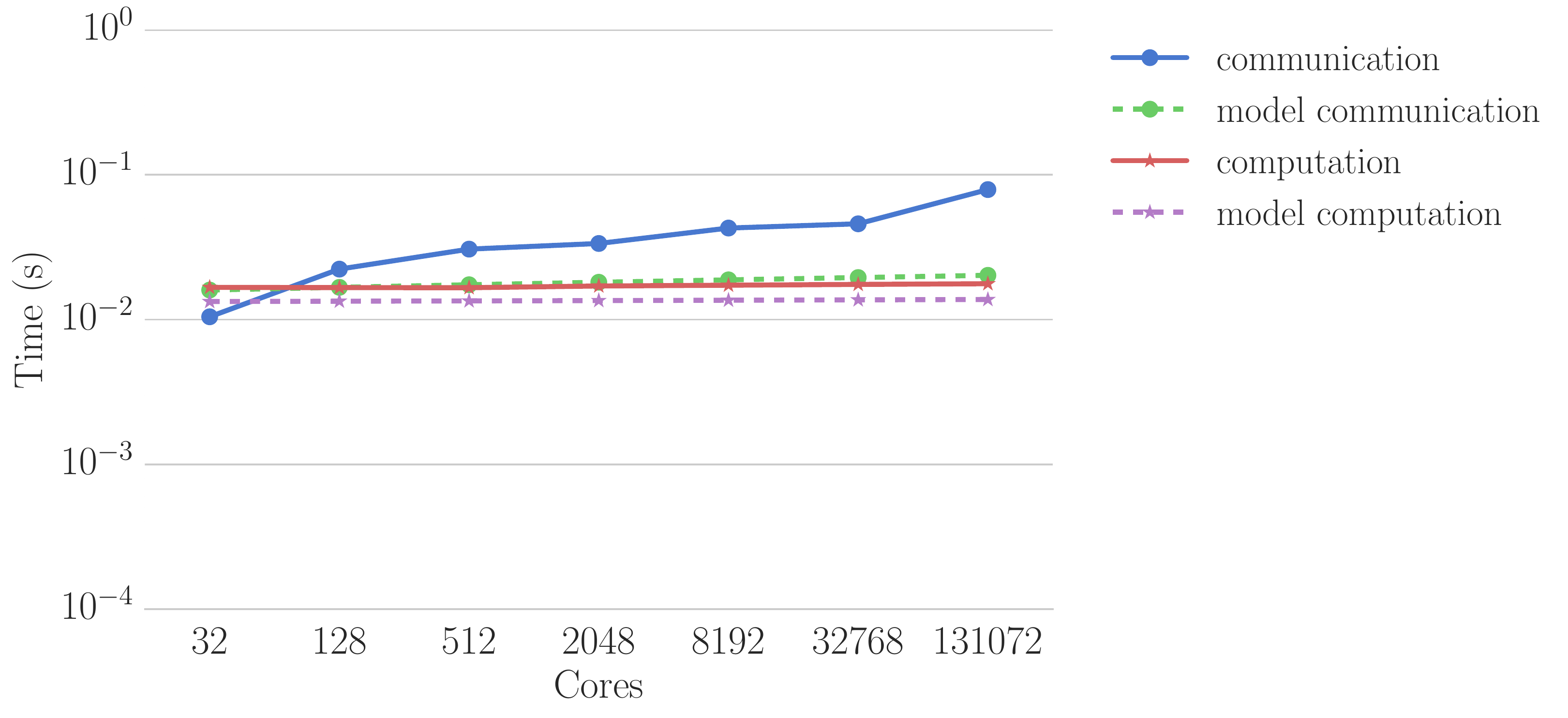}
  \caption{Weak scaling of smoothing routine on Blue Waters with local problem size $568 \times 71$.}\label{fig:bw-ws-split}
\end{figure}

Nevertheless, on Blue Waters communication cost for redistribution
remains relatively low in comparison to smoothing, which remains the
dominant kernel in the overall solve. Figure~\ref{fig:bw-ws-split}
decomposes the cost of smoothing into communication and computation.
This decomposition indicates that the communication cost of the
halo exchange is the primary contributor to the reduced scalability.

In contrast to the network advantages of Mira, lower
floating point performance is observed for the key computational kernels on Mira
than on Blue Waters.  This difference, combined with an extra core
dedicated to operating system functions, yields more predictable
performance and superior scaling behavior.  This is also observed for
a variety of applications~\cite{Kerbyson2012}: loss in performance on
the Cray XE6 in comparison to BG/Q systems as the core count
increases. However, with the data locality and fixed communication patterns of
the algorithm, the network performance on Blue Waters is good enough to
let its superior floating performance yield the best time to solution,
solving approximately 2.5 times faster at 131K cores.

\subsubsection*{Strong Scalability}

\begin{figure}
  \centering
  \includegraphics[width=.9\textwidth]{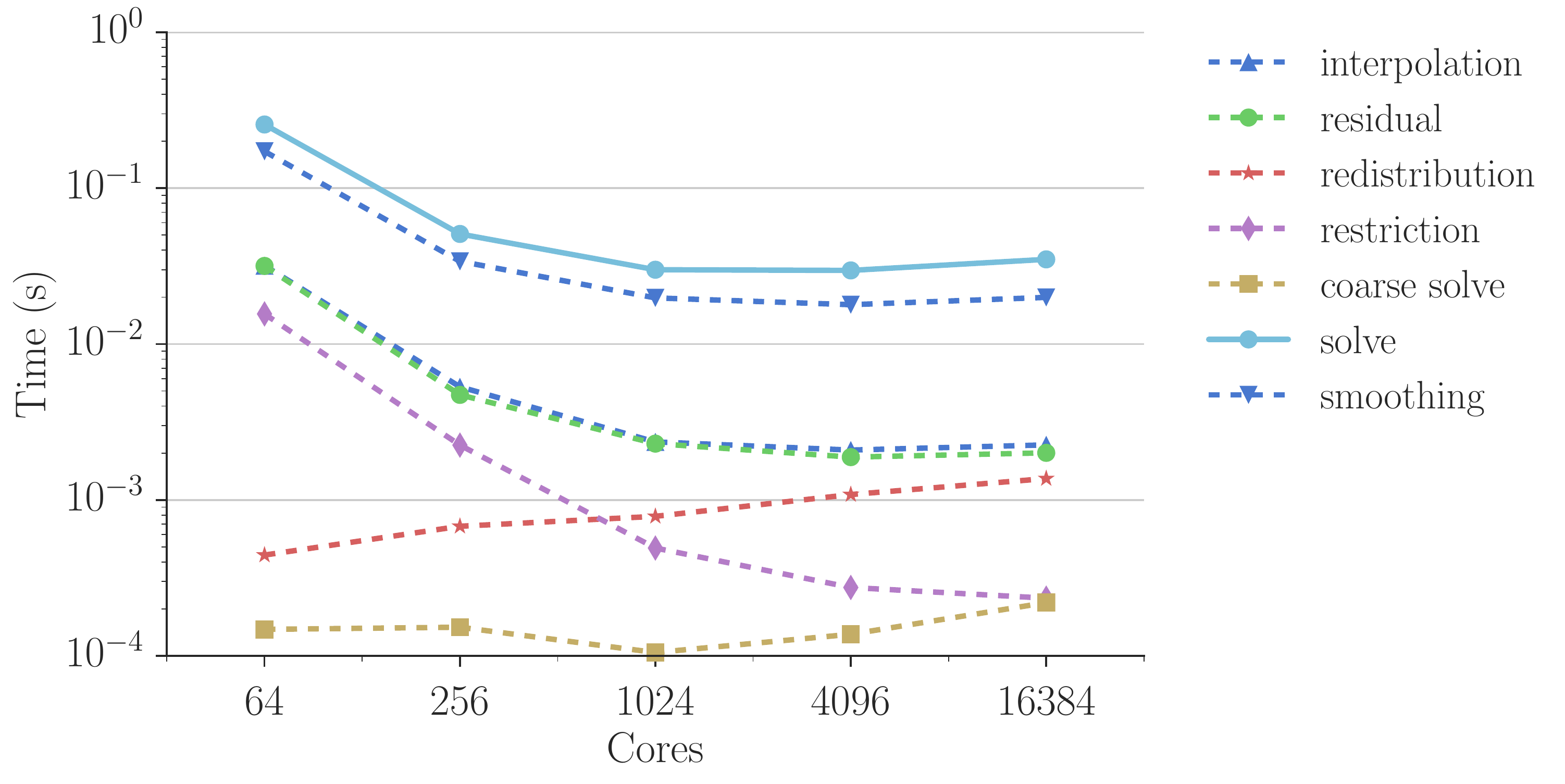}
  \caption{Strong scaling on Blue Waters with problem size: $3200 \times 3200$.}\label{fig:bw-ss}
\end{figure}

Turning to strong scalability, an isotropic model diffusion problem is
setup on a 3200 $\times$ 3200 grid, and then executed on a sequence of
square processor grids.  The number of cores in each coordinate
direction is successively doubled from 8 to 128, to create a range of
core counts from 64 to 16,384, with a corresponding local problem size
ranging from 160,000 to 625 degrees of freedom (dof).  The strong
scaling performance for Blue Waters is shown in
Figure~\ref{fig:bw-ss}.  Here, the strong scaling limit is reached at
1024 cores, which is equivalently 10K dof per core.  This scaling
limit is consistent with other performance assessments of multilevel
solvers in application codes
(e.g.,~\cite{hammond:12,sundar_parallel_2012}).

The computational cost of the solve is again dominated by
smoothing and scaling is limited by communication, as supported in
Figure~\ref{fig:bw-ws-split} by the performance model of
Section~\ref{sec:perf_model}.
Often a hybrid Gauss-Seidel Jacobi sweep is used to limit
communication to once per relaxation sweep~\cite{Baker2012}, improving
the cycles strong scaling even though it may potentially slow
convergence. Here, the focus is on redistribution and
the impact of communication is highlighted with
a strict implementation of four-color Gauss-Seidel,
which results in four halo exchanges per relaxation sweep.
The granularity of the component timers is such that the residual
and interpolation operations each include a single halo exchange.
This leads to strong scaling curves for these operations that
are very similar to smoothing, but much faster in absolute terms.

In contrast, the restriction has no communication, and would exhibit
perfect strong scaling in the absence of redistribution. With
redistribution fewer cores are used at coarser levels, and strong
scaling begins to degrade at 2.5K dof per core, although speedup
is still realized even at 625 dof per core.  This observation
highlights the complexity of trade-offs in multilevel algorithms
on modern systems, and the important role of performance
models in design and runtime optimization.

\begin{figure}
  \centering
  \includegraphics[width=.9\textwidth]{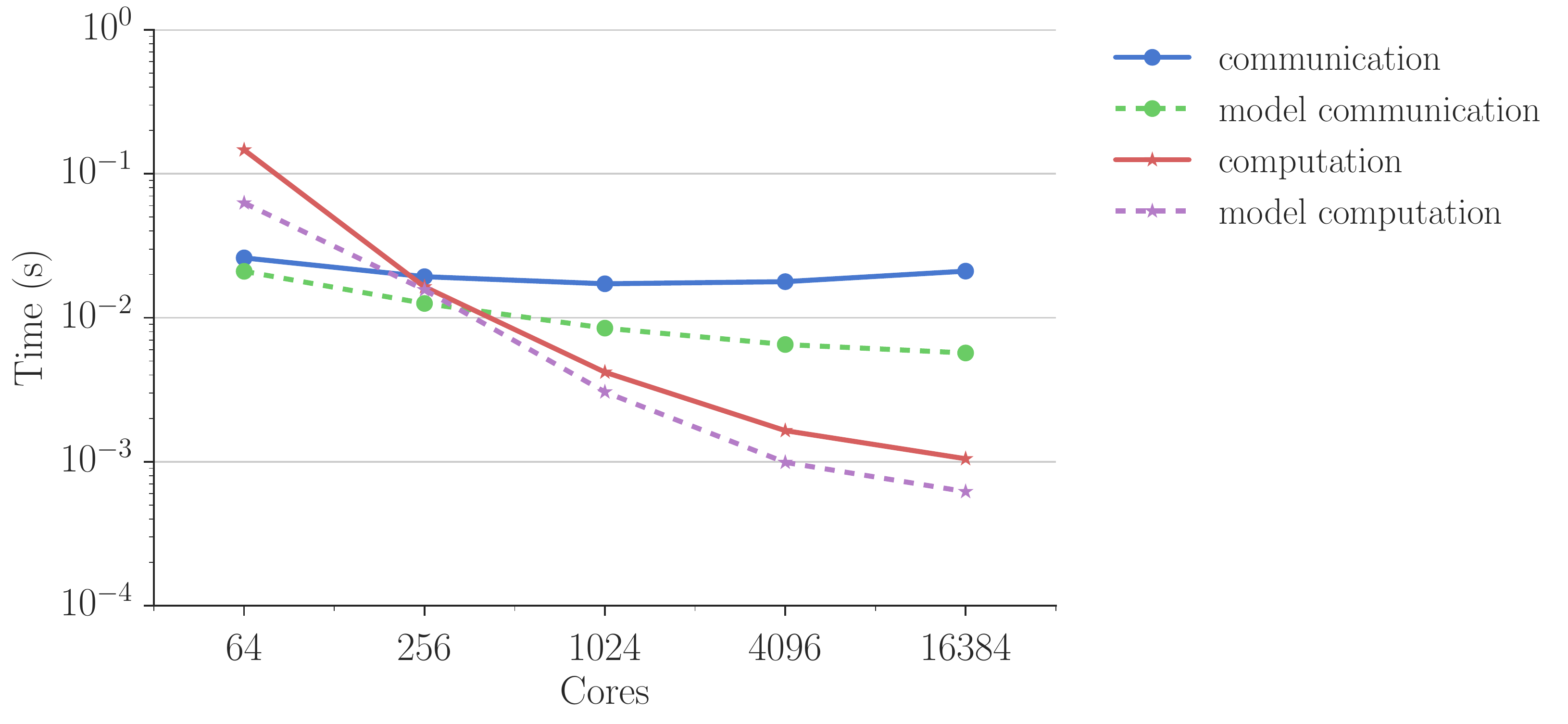}
  \caption{Strong scaling of smoothing routine on Blue Waters with problem size: $3200 \times 3200$.}\label{fig:bw-ss-split}
\end{figure}

\subsection{Performance of Optimized Redistribution}

To understand the impact of the redistribution path on solve time, the
$568 \times 71$ weak scaling problem used in Figure~\ref{fig:bw-40k}
with 2048 processors is executed over a variety of redistribution choices.
The selected redistribution paths are listed in Table~\ref{tbl:paths}.
Path 1 indicates the optimal redistribution path used in
Figure~\ref{fig:bw-40k} and is selected by the algorithm from
Section~\ref{sec:alg}.  In stark contrast to Path 1 is Path 0, which is the
original all-to-one redistribution path that is known to scale poorly.
The remaining paths highlight the flexibility in the agglomeration sequence.
\begin{table}[!htb]
  \setlength{\tabcolsep}{3pt}
  \centering
  \begin{tabular}{cllllllllllllll}
    \toprule
    Path & \multicolumn{14}{c}{Processor Grid Redistribution} \\
    \midrule
    0 & $64 \times 32$ & $\rightarrow$ & $1  \times 1 $  &               &               &               &               &               &               &               &               &               &             \\
    1 & $64 \times 32$ & $\rightarrow$ & $64 \times 16$  & $\rightarrow$ & $64 \times 8$ & $\rightarrow$ & $64 \times 4$ & $\rightarrow$ & $32 \times 2$ & $\rightarrow$ & $16 \times 1$ & $\rightarrow$ & $1 \times 1$\\
    2 & $64 \times 32$ & $\rightarrow$ & $64 \times 4 $  & $\rightarrow$ & $8  \times 1$ & $\rightarrow$ & $4 \times 1$  &               &               &               &               &               &             \\
    3 & $64 \times 32$ & $\rightarrow$ & $16 \times 2 $  & $\rightarrow$ & $1  \times 1$ &               & &             &               &               &               &               &               &             \\
    4 & $64 \times 32$ & $\rightarrow$ & $64 \times 16$  & $\rightarrow$ & $2  \times 1$ & $\rightarrow$ & $1 \times 1$  &               &               &               &               &               &             \\
    5 & $64 \times 32$ & $\rightarrow$ & $4  \times 1 $  & $\rightarrow$ & $1  \times 1$ &               & &             &               &               &               &               &               &             \\
    6 & $64 \times 32$ & $\rightarrow$ & $64 \times 16$  & $\rightarrow$ & $4  \times 1$ & $\rightarrow$ & $1 \times 1$  &               &               &               &               &               &             \\
    7 & $64 \times 32$ & $\rightarrow$ & $64 \times 16$  & $\rightarrow$ & $64 \times 8$ & $\rightarrow$ & $2 \times 1$  & $\rightarrow$ & $1 \times 1$  &               &               &               &             \\
    8 & $64 \times 32$ & $\rightarrow$ & $2  \times 1 $  & $\rightarrow$ & $1  \times 1$ &               &               &               &               &               &               &               &             \\
    \bottomrule
  \end{tabular}
  \caption{Example redistribution paths using a $64 \times 32$ processor grid and $568 \times 71$ local problem size.}\label{tbl:paths}
\end{table}

\begin{figure}[!htb]
  \centering
  \includegraphics[width=.8\textwidth]{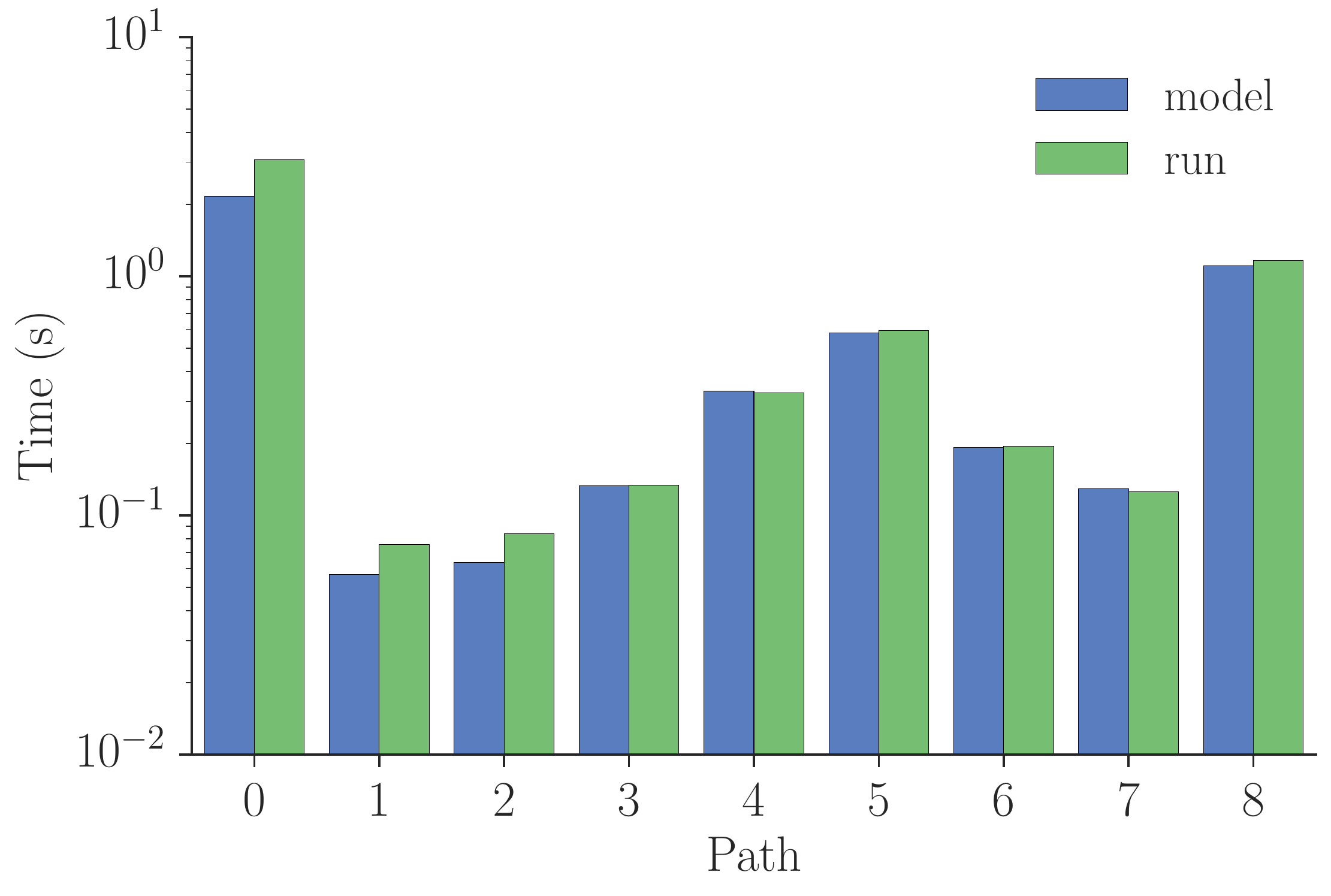}
  \caption{Model and run times on Blue Waters of various redistribution paths using a $64 \times 32$ processor grid and $568 \times 71$ local problem size.}\label{fig:psens}
\end{figure}

The bar graph in Figure~\ref{fig:psens} compares the run times and
predicted times of the multigrid solve phase for the various
redistribution paths shown in Table~\ref{tbl:paths}.  The predicted
times are computed using the performance model from
Section~\ref{sec:perf_model}, and are in good agreement with the
actual run times. Comparing the run times of Path 0 and Path 1, this
graph shows a 40 times speedup of the new redistribution strategy
over the original all-to-one strategy.  In addition, Path 1
has the lowest run time of several plausible alternatives to the
all-to-one strategy.  Thus, Figure~\ref{fig:psens}
demonstrates the utility of using a global search guided by a
performance model as a predictive tool for choosing optimal
redistribution paths.
Moreover, using this strategy to select the
redistribution path delivered effective weak parallel scaling out to 500K+
cores on Mira, and out to 132K+ cores on Blue Waters, while the
original code was essentially unusable beyond 4K cores.

\section{Conclusions}\label{sec:conclusions}
Emerging architectures place an increasing importance on data locality
and minimizing data movement.  In these environments, structured
approaches benefit from predictable memory access patterns and
avoiding indirect addressing.  This motivates the development of
methods that exploit local structure to avoid incurring the performance
consequences of full algebraic generality.  A robust structured single-block
multigrid implementation that scales well on modern architectures is an
important step toward this goal.

In this paper, a new optimized recursive agglomeration algorithm for
redistributing coarse-grid work in structured multilevel solvers is
introduced. This algorithm combines a predictive performance model
with a structure exploiting recursive enumeration of coarse processor
grids to enable a global search for the optimal agglomeration
strategy.  This approach significantly improves the weak parallel
scalability of robust, structured multigrid solvers such as BoxMG\@.
In this study using BoxMG operators, this new algorithm delivers very
good weak scaling up to $524,288$ cores on an IBM BG/Q (Mira), and
reasonable weak scaling up to $131,072$ cores on Cray XE (Blue
Waters).
The speedup over the previous all-to-one agglomeration approach is
significant even at modest core counts, 40 times speedup on just 2048
cores and 144 times speedup on 8192 cores.  At larger core counts, the
all-to-one agglomeration approach became infeasible as the increased
memory requirements for the coarse-grid problem exceeded available
memory.

In addition, the strong scaling of multigrid solves using this
redistribution algorithm is demonstrated on Blue Waters. Overall, the
scaling limit is observed to be approximately 10K dofs per core, which
is similar to results obtained in other studies and is dominated by
the communication cost of the smoother.  Future work on additive
variants of multigrid and improvements to the performance model to
more accurately capture deep memory hierarchies are needed to reduce
this limit.

To focus on coarse-grid redistribution, only pointwise Gauss-Seidel
relaxation is considered in this paper. In the future, smoothers that
may enhance performance, such as hybrid or polynomial smoothers, or
enhance robustness, such as line and plane smoothers
\cite{austin-2004-linesolver}, may be considered and their
impact on optimal coarse-grid redistribution explored.

\section*{Acknowledgments}

This work was carried out under the auspices of the National
Nuclear Security Administration of the U.S. Department of Energy, at
the University of Illinois at Urbana-Champaign under Award Number
DE-NA0002374, and at Los Alamos National Laboratory under Contract
Number DE-AC52-06NA25396, and was partially supported by the Advanced Simulation and
Computing / Advanced Technology Development and Mitigation Program.
This research is part of the Blue Waters sustained-petascale computing
project, which is supported by the National Science Foundation (awards
OCI-0725070 and ACI-1238993) and the state of Illinois. Blue Waters is
a joint effort of the University of Illinois at Urbana-Champaign and
its National Center for Supercomputing Applications.
This research used resources of the Argonne Leadership Computing
Facility, which is a DOE Office of Science User Facility supported
under Contract DE-­‐AC05-­‐00OR22725.

\bibliographystyle{siamplain}
\bibliography{references}
\end{document}